\documentclass[prl,twocolumn,showpacs,preprintnumbers,amsmath,superscriptaddress,amssymb]{revtex4-1}

\usepackage{graphicx}
\usepackage{bm}
\usepackage{color}

\begin{document}



\title{Minimal model for charge transfer excitons at the dielectric interface}
\author{Shota Ono}
\email{shota\_o@gifu-u.ac.jp}
\affiliation{Department of Electrical, Electronic and Computer Engineering, Gifu University, Gifu 501-1193, Japan}
\affiliation{Department of Physics, Graduate School of Engineering, Yokohama National University, Yokohama, 240-8501, Japan}
\author{Kaoru Ohno}
\email{ohno@ynu.ac.jp}
\affiliation{Department of Physics, Graduate School of Engineering, Yokohama National University, Yokohama, 240-8501, Japan}

\begin{abstract}
Theoretical description of the charge transfer (CT) exciton across the donor-acceptor interface without the use of a completely localized hole (or electron) is a challenge in the field of organic solar cells. We calculate the total wavefunction of the CT exciton by solving an effective two-particle Schr\"{o}dinger equation for the inhomogeneous dielectric interface. We formulate the magnitude of the CT and construct a minimal model of the CT exciton under the breakdown of inversion symmetry. We demonstrate that both a light hole mass and a hole localization along the normal to the dielectric interface are crucial to yield the CT exciton.
\end{abstract}




\maketitle

Charge transfer (CT) exciton is a key to resolve a long-standing exciton dissociation problem in organic solar cells \cite{zhu,few,bassler}. Although several effects such as interface dipole \cite{arkhipov,wiemer,yost}, disorder \cite{peumans,vithanage,ferguson}, carrier delocalization \cite{deibel,nenashev}, effective mass \cite{schwarz}, and entropy \cite{clarke,gregg,gao,monahan} have been investigated to understand why the CT exciton dissociates into free carriers efficiently at the dielectric interface, a physics behind the dissociation remains under debate. The CT exciton has been modeled as a pair of a mobile electron in the acceptor and a completely localized hole in the donor or vice versa \cite{arkhipov,wiemer,nenashev,schwarz,monahan,muntwiler,yang}. However, such a treatment partly ignores the spatial correlation between carriers \cite{zhu}. To compute the total wave function of the CT exciton is a theoretical challenge. Raos {\it et al}. proposed an exciton tight-binding model without using the localized particle approximation \cite{raos}. In contrast, a natural extension of the standard hydrogen model would also be useful to understand an origin of the CT.

In this paper, we construct a minimal model of the CT exciton described by a two-particle Schr\"{o}dinger equation for an inhomogeneous system having the local dielectric constant $\varepsilon (\bm{r})$ ($\bm{r}$ is the position). One of the main results is that the carrier localization along the {\it normal} to the dielectric interface enhances the magnitude of the CT. This is complementary to the previous finding that the carrier localization {\it parallel} to the interface lowers the exciton dissociation probability \cite{bassler,nenashev}.

We first present a two-particle Schr\"{o}dinger equation for an inhomogeneous dielectric medium ${\cal H}\psi(\bm{r}_e;\bm{r}_h) = {\cal E} \psi(\bm{r}_e;\bm{r}_h)$, where $\bm{r}_e$ and $\bm{r}_h$ are the electron and hole position, respectively. The Hamiltonian is given by
\begin{eqnarray}
 {\cal H} = -\frac{\hbar^2}{2m_e} \nabla_{e}^{2} -\frac{\hbar^2}{2m_h} \nabla_{h}^{2} 
 + U(\bm{r}_e;\bm{r}_h),
 \label{eq:schrodinger}
\end{eqnarray}
where $m_e$ and $m_h$ are the electron and hole masses, respectively. The first and second terms are the kinetic energies of the electron and hole, respectively. The third term is the Coulomb potential energy between the electron and the hole, which may be decomposed into three terms:
\begin{eqnarray}
 U(\bm{r}_e;\bm{r}_h) 
&=& \frac{-e^2}{4\pi \sqrt{\varepsilon (\bm{r}_e) \varepsilon (\bm{r}_h)} }
G (\bm{r}_e;\bm{r}_h) 
\nonumber\\
&+& V_{\rm im}(\bm{r}_e) + V_{\rm im}(\bm{r}_h),
\label{eq:potential}
\end{eqnarray}
where $e$ is the elementary charge. The first term is the direct interaction energy between particles at $\bm{r}_e$ and $\bm{r}_h$ and the second and third terms are the image potential (IP) energies of particles at $\bm{r}_e$ and $\bm{r}_h$, respectively. The set of Eqs.~(\ref{eq:schrodinger}) and (\ref{eq:potential}) is a general expression describing the two-particle kinetics in any dielectric. Below, we will derive Eq.~(\ref{eq:potential}).

Let us consider an electrostatic potential acting on $\bm{r}$ caused by a creation of a point source charge $Q_s$ placed at the position $\bm{r}_s$. Such a potential $\phi (\bm{r};\bm{r}_s)$ is determined by solving the Poisson equation
\begin{eqnarray}
 \nabla \cdot \left[\varepsilon (\bm{r}) \nabla \phi (\bm{r};\bm{r}_s) \right]= - Q_s \delta (\bm{r} - \bm{r}_s).
 \label{eq:1}
\end{eqnarray}
By noting the following relation
\begin{eqnarray}
 \nabla \cdot \left[\varepsilon (\bm{r}) \nabla \phi (\bm{r};\bm{r}_s) \right] 
&=& \sqrt{\varepsilon (\bm{r})}  \nabla^2 
\left[ \sqrt{\varepsilon (\bm{r})} \phi (\bm{r};\bm{r}_s) \right] \nonumber\\
&-& \sqrt{\varepsilon (\bm{r})}  \left( \nabla^2  \sqrt{\varepsilon (\bm{r})} \right) \phi (\bm{r};\bm{r}_s),
\end{eqnarray}
one can rewrite the Poisson equation as follows
\begin{eqnarray}
 & &\nabla^2 \left[ \sqrt{\varepsilon (\bm{r})} \phi (\bm{r};\bm{r}_s) \right] 
 \nonumber\\
 &=& - \frac{Q_s \delta(\bm{r}-\bm{r}_s)}{\sqrt{\varepsilon (\bm{r}_s)}}
+ \left( \nabla^2  \sqrt{\varepsilon (\bm{r})} \right) \phi (\bm{r};\bm{r}_s).
\label{eq:deform}
\end{eqnarray}
If we regard two terms on the right hand side (rhs) of Eq.~(\ref{eq:deform}) as a source charge for the potential $\sqrt{\varepsilon (\bm{r})} \phi (\bm{r};\bm{r}_s)$, we obtain a self-consistent equation
\begin{eqnarray}
\phi (\bm{r};\bm{r}_s) &=& \Phi_0 (\bm{r};\bm{r}_s) \nonumber\\
 &-& \frac{1}{4\pi  \sqrt{\varepsilon (\bm{r})} }
 \int \frac{ \nabla^2  \sqrt{\varepsilon (\bm{r}')}}{\vert \bm{r} - \bm{r}' \vert}
 \phi (\bm{r}';\bm{r}_s) d\bm{r}',
 \label{eq:poisson_self}
\end{eqnarray}
where
\begin{eqnarray}
\Phi_0 (\bm{r};\bm{r}_s) = \frac{Q_s}{4\pi \sqrt{\varepsilon (\bm{r}) \varepsilon (\bm{r}_s)} }
G_0 (\bm{r}-\bm{r}_s)
\label{eq:phi_0}
\end{eqnarray}
and $G_0 (\bm{r}-\bm{r}_s) \equiv 1/\vert \bm{r} - \bm{r}_s\vert$. This is simply written as
\begin{eqnarray}
\phi (\bm{r};\bm{r}_s)
=\frac{Q_s}{4\pi \sqrt{\varepsilon (\bm{r}) \varepsilon (\bm{r}_s)} }
G (\bm{r};\bm{r}_s),
\label{eq:phidef}
\end{eqnarray}
where
\begin{eqnarray} 
 G (\bm{r};\bm{r}_s) = G_0 (\bm{r}-\bm{r}_s) 
+\int d\bm{r}' G_0 (\bm{r}-\bm{r}') p(\bm{r}') G (\bm{r}';\bm{r}_s),
\nonumber\\
\label{eq:Gdef}
\end{eqnarray}
and $p(\bm{r}) = - \nabla^2 \sqrt{\varepsilon (\bm{r})}/[4\pi \sqrt{\varepsilon(\bm{r})}]$. The electrostatic potential energy between the point charge $Q$ at the position $\bm{r}$ and the source charge $Q_s$ at the position $\bm{r}_s$ is given as $Q \phi (\bm{r};\bm{r}_s)$. The second term on the rhs of Eq.~(\ref{eq:Gdef}) contributes to the induced potential caused by the presence of the spatially varying $\varepsilon$. The limit $\bm{r}\rightarrow\bm{r}_s$ of the induced potential yields the IP \cite{granger,segui}
\begin{eqnarray}
 V_{\rm im} (\bm{r}_s) 
&=&\frac{Q_{s}^{2}}{8\pi \varepsilon (\bm{r}_s) }
\left[ G (\bm{r}_s;\bm{r}_s) - G_0 (\bm{r}_s - \bm{r}_s) \right].
\label{eq:vim}
\end{eqnarray}
 Consequently, by setting $Q=-Q_s=-e$, $\bm{r}=\bm{r}_e$, and $\bm{r}_s=\bm{r}_h$, we obtain the potential energy, i.e., Eq. (\ref{eq:potential}).

The derivation of Eqs.~(\ref{eq:schrodinger}) and (\ref{eq:potential}) paves the way to study two-particle properties in inhomogeneous dielectric media. All the two-particle problems can be reduced (i) to construct a model of $\varepsilon (\bm{r})$ that captures the underlying physics and (ii) to solve the two-particle Schr\"{o}dinger equation. The locality of $\varepsilon (\bm{r})$ is justified when the length scales in question exceed the spread of maximally localized Wannier functions \cite{giustino,giustino2}. The present work does not use {\it ab initio} approach to determine $\varepsilon(\bm{r})$, but rather use a model to simplify the physics.

\begin{figure}[t]
\center
\includegraphics[scale=0.4,clip]{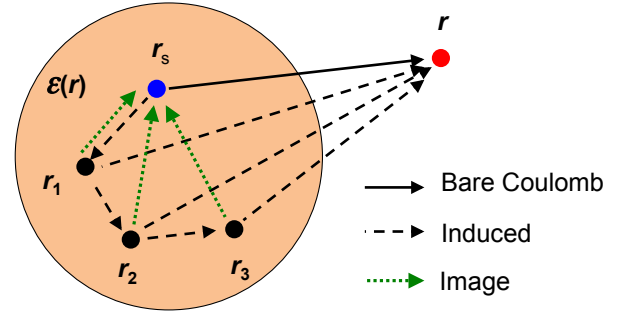}
\caption{\label{fig:born} (Color online) Schematic illustration for the electrostatic potential at $\bm{r}$ by the presence of the source charge at $\bm{r}_s$. Solid and dashed lines indicate the contribution from the bare Coulomb and induced potential, respectively. The dotted line indicates the IP contribution to the position $\bm{r}_s$.}
\end{figure}

For a homogeneous system ($p\rightarrow 0$), Eq.~(\ref{eq:schrodinger}) is clearly reduced to a standard effective mass equation for bulk semiconductors. For an inhomogeneous system, both the center of mass motion and the relative motion cannot be separated because $\phi (\bm{r};\bm{r}_s)$ in Eq.~(\ref{eq:phidef}) is no longer a function of $\bm{r}-\bm{r}_s$. To explore the physical meaning of Eq.~(\ref{eq:phidef}), we rewrite it as follows
\begin{eqnarray}
 \phi (\bm{r};\bm{r}_s) = \Phi_0 (\bm{r};\bm{r}_s) 
 + \sum_{n=1}^{\infty} \phi^{(n)} (\bm{r};\bm{r}_s),
\end{eqnarray}
where
\begin{eqnarray}
 \phi^{(n)} (\bm{r};\bm{r}_s) = \int d\bm{r}_{n}
\frac{\rho^{(n)}(\bm{r}_n;\bm{r}_s)}
{4\pi \sqrt{\varepsilon (\bm{r})\varepsilon (\bm{r}_n)}} G_0 (\bm{r} - \bm{r}_n)
\label{eq:phi_n}
\end{eqnarray}
with $n$-th ($n\ge 1$) order induced charge
\begin{eqnarray}
& &\rho^{(n)}(\bm{r}_n;\bm{r}_s)
 =  4\pi \varepsilon (\bm{r}_n) p (\bm{r}_n) \nonumber\\
 &\times&
 \int d\bm{r}_{n-1} 
\frac{\rho^{(n-1)}(\bm{r}_{n-1};\bm{r}_s)}
{4\pi \sqrt{\varepsilon (\bm{r}_n)\varepsilon (\bm{r}_{n-1})}} G_0 (\bm{r}_n - \bm{r}_{n-1})
\label{eq:rho_n}
\end{eqnarray}
and $\rho^{(0)}(\bm{r}_0;\bm{r}_s) = Q_s \delta (\bm{r}_0-\bm{r}_s)$. Figure~\ref{fig:born} indicates how a charge at $\bm{r}$ interacts with a source charge at $\bm{r}_s$ via the induced potential (dashed) as well as the bare Coulomb potential (solid): the $(n-1)$-th ($n\ge 1$) order charge density $\rho^{(n-1)}$ at $\bm{r}_{n-1}$ produces the $n$th-order induced charge density $\rho^{(n)}$ at $\bm{r}_n$ given by Eq.~(\ref{eq:rho_n}), which yields the $n$th-order induced potential $\phi^{(n)}$ at $\bm{r}$ given by Eq.~(\ref{eq:phi_n}). Similar interpretation can also be applied to the physical meaning of IP in Eq.~(\ref{eq:vim}). All these treatments are exact in static electrodynamics. Although the time-evolution of the induced charge and/or potential can be studied in the framework of the linear response theory \cite{schone,gumhalter, cui, ono, gumhalter2}, such a problem is out of scope in this paper.


\begin{figure*}[t]
\center
\includegraphics[scale=0.8,clip]{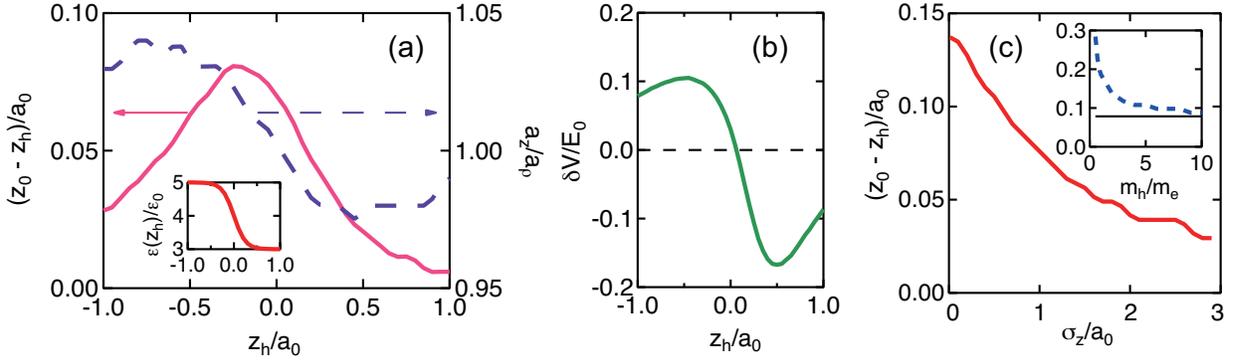}
\caption{\label{fig:ct} (Color online) (a) $z_h$-dependence both of the values of $z_0-z_h$ (left) and $a_z/a_\rho$ (right) for the dielectric interface. The trapped hole approximation is assumed: $m_h/m_e\rightarrow \infty$ and $\sigma_z \rightarrow 0$. Inset: the spatial profile of $\varepsilon$ in Eq.~(\ref{eq:model_ct}). (b) $\delta V$ in Eq.~(\ref{eq:deltaV}) as a function of $z_h$. (c) The charge transfer $z_0-z_h$ as a function of $\sigma_z$ for $z_h = -0.2 a_0$ and $m_h/m_e = 2$. Inset: The charge transfer $z_0-z_h$ as a function of $m_h/m_e$ for $z_h = -0.2 a_0$ and $\sigma_z = 0.01a_0$ (see Eq.~(\ref{eq:trial1})). The thin solid line indicates the CT value in the limit of $m_h/m_e\rightarrow \infty$.}
\end{figure*}

Based on the formulation above, we next study the CT exciton problem. We consider a dielectric interface, in which the dielectric constant varies only along the normal to the interface \cite{xue}. The magnitude of the dielectric constant is given by
\begin{eqnarray}
 \varepsilon(z) = \frac{\varepsilon_{\rm in} + \varepsilon_{\rm out}}{2}
 - \left( \frac{\varepsilon_{\rm in} - \varepsilon_{\rm out}}{2} \right) \tanh \left( \frac{z}{w}\right),
 \label{eq:model_ct}
\end{eqnarray}
where $\varepsilon_{\rm in}$ and $\varepsilon_{\rm out}$ are the bulk dielectric constant in inner and outer regions, respectively: $\lim_{z\rightarrow -\infty(+\infty)}\varepsilon(z) = \varepsilon_{\rm in(out)}$. The parameter $w$ determines the smoothness of the dielectric constant variation near $z=0$: The limit $w\rightarrow 0$ gives a step function. The density-functional theory approach has shown that the dielectric constant changes monotonically around the semiconductor interface, while a slight deviation from the monotonic curve appears due to the presence of the atomic nuclei but vanishes in each bulk region \cite{giustino2}. The use of Eq. (\ref{eq:model_ct}) would be enough to construct a CT exciton minimal model. In this case study, we set $\varepsilon_{\rm in}=5\varepsilon_0$, $\varepsilon_{\rm out}=3\varepsilon_0$ ($\varepsilon_0$ is the dielectric constant of vacuum), which is typical values of organic semiconductors, and $w=a_0/4$ that corresponds to the transition region width of $a_{\rm 0} = 4\pi \varepsilon_{\rm in}\hbar^2/(m_0 e^2) (\simeq 2.64$\ \AA) around $z=0$ (see the inset of Fig.~\ref{fig:ct}(a)). The energy unit is set to be $E_0 = e^2/(8\pi \varepsilon_{\rm in} a_{\rm 0})= 1/25$ Ry. We set $m_e=0.8m_0$ by referring to the electron mass of pentacene \cite{muntwiler,yang}.

\begin{figure}[t]
\center
\includegraphics[scale=0.45,clip]{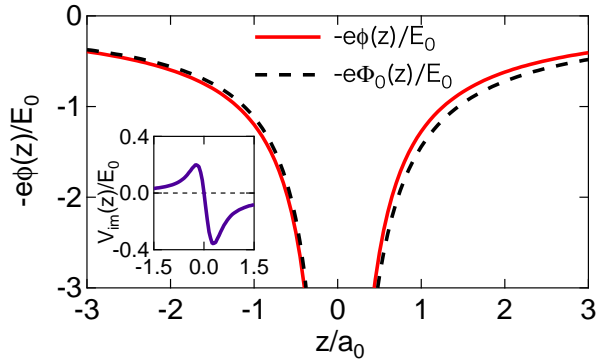}
\caption{\label{fig:cou} (Color online) Self-consistent (solid) and zero-th order (dashed) solution to the Poisson equation given by Eq.~(\ref{eq:poisson_self}). Both solution is asymmetric with respect to $z=0$. Inset: The IP energy given by Eq.~(\ref{eq:vim}). }
\end{figure}

Figure \ref{fig:cou} shows the $z$-dependence of $\Phi_0$ and $\phi$ in the presence of a hole at $z_h = 0$: the former is simply given by Eq.~(\ref{eq:phi_0}) with a replacement of $Q_s$ by $+e$, while the latter are obtained by solving Eq.~(\ref{eq:poisson_self}) self-consistently (i.e., exactly). Due to the small difference between $\varepsilon_{\rm in}$ and $\varepsilon_{\rm out}$, the potential energy difference between them is also small. This fact also holds for other hole position $z_h$. The inset of Fig.~\ref{fig:cou} shows the $z$-dependence of IP given by Eq.~(\ref{eq:vim}). As a charged particle approaches the interface from outside, $V_{\rm im} (z)$ decreases in the vicinity of the interface, increases in the transition region, and takes a maximum at $z\simeq -0.25 a_0$. In the limit of $z\rightarrow -\infty$, $V_{\rm im} (z)$ approaches zero. Similar spatial variation of the IP has been reported in a semiconductor interface model \cite{xue}. The IP around a metal-vacuum interface has been studied by a fully quantum mechanical method \cite{gumhalter2}. The IP variation for $z>0$ is also similar to that in the vacuum region at the metal-vacuum interface. This fact supports the validity of the present model. In this study, $\phi- \Phi_0$ and $V_{\rm im} (z)$ are small enough to be neglected in the first approximation.


Motivated by the result above, we retain the zero-th order potential $\Phi_0$ given in Eq.~(\ref{eq:phi_0}) only for the Coulomb interaction between an electron at $\bm{r}_e$ and a hole at $\bm{r}_h$. This yields the two-particle Hamiltonian
\begin{eqnarray}
 {\cal H} = -\frac{\hbar^2}{2M} \nabla_{\bm{R}}^{2}  -\frac{\hbar^2}{2\mu} \nabla_{\tilde{\bm{r}}}^{2}  
 -e\Phi_0\left(\bm{R} + \frac{m_h}{M} \tilde{\bm{r}};\bm{R} - \frac{m_e}{M} \tilde{\bm{r}}\right),
 \nonumber\\
\end{eqnarray}
where $M = m_e + m_h$ and $\mu = m_e m_h/M$ are the total and reduced masses, respectively. $\bm{R} = (m_e\bm{r}_e+m_h\bm{r}_h)/M$ and $\tilde{\bm{r}} = \bm{r}_e-\bm{r}_h$ are the center-of-mass and the relative coordinates, respectively. 

To solve the two-particle Schr\"{o}dinger equation ${\cal H}\Psi (\bm{R};\tilde{\bm{r}}) = {\cal E}\Psi (\bm{R};\tilde{\bm{r}})$, we use the variational approach and define the ground state trial function as
\begin{eqnarray}
 & & \Psi (\bm{R};\tilde{\bm{r}}; a_{\rho}, a_{z}, z_0) 
 = \frac{1}{(2\pi)^{3/2}} e^{i(K_x X + K_y Y)}e^{im\tilde{\theta}}
 \nonumber\\
 &\times&
 \xi (Z; \tilde{z}) \psi (\tilde{r}, \tilde{z}; a_{\rho}, a_{z}, z_0)
 \label{eq:trial}
\end{eqnarray}
with
\begin{eqnarray}
& & \xi (Z;\tilde{z}) = \frac{1}{(\pi \sigma_{z}^{2})^{1/4}} 
\exp{\left[-\frac{1}{2} \left( \frac{Z - \frac{m_e}{M}\tilde{z} - z_{h}}{\sigma_{z}}\right)^2 \right]}
\label{eq:trial1}
\end{eqnarray}
and
\begin{eqnarray} 
& & \psi (\tilde{r}, \tilde{z};a_{\rho}, a_{z}, z_0) 
\nonumber\\
 &=&\frac{1}{\sqrt{{\pi a_{\rho}^{2} a_{z}}}} 
 \exp \left(- \sqrt{\left(\frac{\tilde{\rho}}{a_{\rho}}\right)^2 
 + \left(\frac{\tilde{z}-z_0}{a_{z}}\right)^2} \right).
 \label{eq:trial2}
\end{eqnarray} 
We used the Cartesian coordinates $\bm{R}=(X,Y,Z)$ and cylindrical coordinates $\tilde{\bm{r}}=(\tilde{\rho}, \tilde{\theta}, \tilde{z})$ for the center-of-mass and the relative coordinates, respectively. Due to the homogeneity parallel to the $xy$-plane and the rotational symmetry around the $z$-axis, the wavefunction is characterized by the wavenumbers $K_x$ and $K_y$ and the angular momentum $m$, respectively (see Eq.~(\ref{eq:trial})). In the following, we set $m = 0$. We assumed that the hole amplitude has a gaussian distribution which is localized at $z = z_h$ and has an extent of $\sigma_z$ along the $z$-direction (see Eq.~(\ref{eq:trial1})). The function $\psi$ in Eq.~(\ref{eq:trial2}) has three variational parameters: $a_{\rho}$ and $a_{z}$ determine an extent of the electronic wavefunction along the $\tilde{\rho}$- and $\tilde{z}$-direction, respectively; $z_0$ determines $z$-coordinate of the center of the electron density distribution. This trial function satisfies the normalization condition
\begin{eqnarray}
 \int d\bm{R} \int d\tilde{\bm{r}}
  \left\vert \Psi (\bm{R};\tilde{\bm{r}}; a_{\rho}, a_{z}, z_0)   \right\vert^2 = 1.
\end{eqnarray}

When the values of $m_e$, $m_h$, $z_h$, and $\sigma_z$ are given, the equation that should be solved is explicitly written as
\begin{eqnarray}
 \left[ -\frac{\hbar^2}{2\mu} \nabla_{\tilde{\bm{r}}}^{2} 
  - \frac{e^2}{4\pi\vert \tilde{\bm{r}}\vert}
   \int   \Xi (Z;\tilde{z}) dZ
 \right]
 \psi (\tilde{\bm{r}}) 
 = E \psi (\tilde{\bm{r}})
 \label{eq:two_org}
\end{eqnarray}
with 
\begin{eqnarray}
 \Xi (Z;\tilde{z})  = \frac{\vert \xi (Z;\tilde{z}) \vert^2}
  {\sqrt{\varepsilon(Z + \frac{m_h}{M}\tilde{z}) \varepsilon(Z - \frac{m_e}{M}\tilde{z})}},
\end{eqnarray}
where $E = {\cal E} - \hbar^2 (K_{x}^{2} + K_{y}^{2})/(2M) - E_{\rm loc}$ with $E_{\rm loc} = \hbar^2 /(4M\sigma_{z}^{2}) + \hbar^2 m_{e}^{2}/(4\mu M^2\sigma_{z}^{2})$ that arises from the hole localization. Note that the standard virial theorem $-V/T=2$, where $T$ and $V$ are the expectation values of the first and second terms in the bracket in Eq.~(\ref{eq:two_org}), respectively, is not satisfied at the dielectric interface because $\Phi_0$ is no longer a function of $\bm{r}_e-\bm{r}_h$ as mentioned. Instead, the relation $-(V + \delta V)/T=2$ should be satisfied where $\delta V$ is defined as
\begin{eqnarray}
 \delta V = \int d\tilde{\bm{r}} 
  \frac{e^2 \vert \psi (\tilde{\bm{r}})\vert^2}{4\pi \vert \tilde{\bm{r}}\vert} 
  \lim_{L\rightarrow 1} 
  \left[\frac{\partial \Xi(L Z;L\tilde{z})}{\partial L} 
  \right].
  \label{eq:deltaV}
\end{eqnarray}
Here $L$ is a scaling parameter. The use of Eq.~(\ref{eq:trial}) gives a ratio of $-(V + \delta V)/T=2.00\pm 0.01$ for the considered systems.

We first consider a hole as a trapped particle. This corresponds to taking both the limit of an infinite hole mass $m_h \rightarrow \infty$ and a strong localization $\sigma_z \rightarrow 0$, which leads to $M\rightarrow \infty$, $\mu \rightarrow m_e$, and $\bm{R} \rightarrow \bm{r}_h$. This treatment may be valid if one of the two phases is disordered \cite{peumans,vithanage,ferguson,deibel,nenashev}. Using Eq.~(\ref{eq:phi_0}), we obtain
\begin{eqnarray}
  \left[ -\frac{\hbar^2}{2m_e} \nabla_{\tilde{\bm{r}}}^{2}  - 
 e\Phi_0\left(\bm{r}_h + \tilde{\bm{r}};\bm{r}_h \right)\right]
 \psi (\tilde{\bm{r}}) = E \psi (\tilde{\bm{r}}),
 \label{eq:two}
\end{eqnarray}
the solution of which gives the binding energy. Figure \ref{fig:ct}(a) (left) shows $z_{h}$ dependence of $z_0 - z_h$. As the hole approaches the interface from the region with $\varepsilon_{\rm in}$ to the region with $\varepsilon_{\rm out}$, $z_0 - z_h$ first increases and reaches its maximum value at $z_h \simeq -0.2 a_0$. Then it gradually decreases within the interface region and goes to zero. The deviation of $z_0 - z_h$ from zero indicates, by definition, the CT exciton. Figure \ref{fig:ct}(a) (right) shows a ratio $a_z/a_{\rho}$ as a function of $z_h$. The ratio deviates from unity largely around the edge of the transition region ($z_h \simeq \pm 0.5 a_0$), indicating the presence of the elongated ($z_h <0$) and shortened ($z_h > 0$) exciton along the $z$-direction and implying that the anisotropy is a precursor of the CT exciton. The behavior in $z_h>0$ can be understood as follows: If the exciton is elongated along the $z$-direction, the potential energy gain decreases due to the high value of $\varepsilon (z)$ in the region of $z<0$. Thus, this leads to a shrinkage of the exciton along the $z$-direction. In contrast, such a shrinkage enhances an exciton kinetic energy, which in turn enhances the exciton extent along the $\rho$-direction to compensate for the kinetic energy loss with the potential energy. Interestingly, we found that the $z_h$-dependence of $\delta V$ in Eq.~(\ref{eq:deltaV}) is strongly correlated with that of $a_z/a_\rho -1$ (see Figs.~\ref{fig:ct}(a) and \ref{fig:ct} (b)). Note that the virial theorem in the present system is given by $2T + V = - \delta V$. Thus, it is reasonable to interpret $- \delta V$ as an effective pressure exerted on the system along the normal to the interface: if $\delta V$ is negative (positive), the effective pressure arising from the inhomogeneity of $\varepsilon (z)$ shortens (elongates) the electron distribution along the $z$ direction. Our model shows that the inversion symmetry breakdown yields a finite value of CT and serves as a minimal model of the CT exciton.

Next, we study the effects of both the finite hole mass and the hole delocalization and solve the eigenvalue problem given by Eq.~(\ref{eq:two_org}). The inset of Fig.~\ref{fig:ct}(c) shows the $m_h/m_e$-dependence of $z_0 - z_h$ in the case of $z_h = -0.2a_0$ and $\sigma_z = 0.01a_0$. The CT value $z_0 - z_h$ monotonically increases with decreasing $m_h/m_e$. The decrease in the hole mass leads to a decrease in the relative mass, $\mu$, which causes a decrease in the binding energy and an increase in the exciton size in real space. The magnitude of the CT would be enhanced in such a weakly bound exciton. Similar behavior has also been reported in other models considering a completely localized carriers \cite{arkhipov,wiemer,nenashev}. Figure \ref{fig:ct}(c) shows the $\sigma_z$-dependence of $z_0 - z_h$ in the case of $m_h/m_e = 2$. As $\sigma_z$ increases (the hole becomes delocalized), the magnitude of the CT decreases. This is because the delocalization of the hole along the normal to the interface leads to a localization of the electron to gain the attractive Coulomb interaction energy, which leads to generation of a tightly bound exciton and thus a decrease in the magnitude of the CT. Our result suggests that the carrier localization normal to the interface would be another key to the exciton dissociation, while it has been suggested that the carrier delocalization parallel to the interface enhances the dissociation probability \cite{nenashev}. More investigation about the relation between the carrier distribution and the interface morphology \cite{jackson} is desired.


In summary, we have studied the total wavefunction of the CT exciton and found that the effects of the inversion symmetry breakdown, the small ratio of $m_h/m_e$, and the hole localization are important in the CT exciton generation. In particular, we expect that experiments can demonstrate the dissociation probability enhancement by the carrier localization along the normal to the interface. 

\begin{acknowledgments}
This study is supported by a Grant-in-Aid for Young Scientists B (No. 15K17435) from JSPS.
\end{acknowledgments}


\end{document}